\documentclass[
aps,
prapplied,
reprint,
superscriptaddress,
amsmath,
amssymb,
floatfix
]{revtex4-2}

\usepackage{graphicx}
\usepackage[colorlinks=true, citecolor=blue]{hyperref}

\begin{document}

\title{Band Gaps and Localization of Surface Waves at Hyperbolic Material and Topological Insulator Interfaces}

\author{Andrei I. Maimistov}
\thanks{Deceased.}
\affiliation{National Nuclear Research University MEPhI, Moscow, 115409 Russia}

\author{Ildar R. Gabitov}
\affiliation{Department of Mathematics, University of Arizona, Tucson, AZ 85721, USA}

\author{Ilia Kuk}
\email{Contact author: ilyakuk@arizona.edu}
\affiliation{Department of Mathematics, University of Arizona, Tucson, AZ 85721, USA}

\date{\today}

\begin{abstract}
	We develop analytical criteria for surface electromagnetic waves supported at the interface between a hyperbolic material and a topological insulator. The analysis identifies the conditions for wave propagation, field penetration into the adjacent media, and the formation of spectral band gaps. In the lossless model, the band edge is governed primarily by the dielectric contrast, while the topological magnetoelectric coupling produces only a small shift of the band edge and a weak correction to the dispersion. For structures combining hexagonal boron nitride with bismuth selenide, the predicted propagation window is narrow when the topological insulator is treated as a bulk medium. A simple effective medium estimate for a thin film suggests that placing a thin topological insulator film on a dielectric substrate with low permittivity could substantially broaden this window. We also analyze a dissipative hyperbolic effective medium composed of titanium and silicon and interfaced with bismuth selenide. In this lossy system, satisfying the dispersion relation is not sufficient to identify a physically admissible guided mode. The fields must also decay away from the interface, attenuate consistently along the propagation direction, have finite penetration depths, and exhibit a consistent energy flow profile. The calculations further show that the orientation of the optical axis strongly affects confinement and attenuation, while apparent discontinuities in a tracked dispersion branch can arise from branch changes in the complex transverse decay constants. These results provide design guidelines for controlling the dispersion, confinement, and propagation loss of surface waves in mid-infrared photonic structures.
\end{abstract}

\maketitle

\section{Introduction}
\label{sec:introduction}

Topological insulators in three spatial dimensions are materials with an insulating bulk and conducting surface states protected by time reversal symmetry~\cite{hasan2010colloquium,qi2011topological}. Their electromagnetic response can be described by axion electrodynamics, where a topological contribution governed by $\theta$ modifies Maxwell equations at interfaces where $\theta$ changes discontinuously~\cite{qi2008topological,wilczek1987two}. In this description, $\theta$ is a dimensionless axion angle, while the strength of the magnetoelectric response is set by the fine-structure constant
$\alpha_{\mathrm{fs}}=e^2/\hbar c$. At an interface, only the jump
$\Delta\theta=\theta_2-\theta_1$ is observable, and the corresponding dimensionless interface coupling is
\begin{equation}
	\alpha
	=
	\alpha_{\mathrm{fs}}
	\frac{\Delta\theta}{\pi}.
	\label{eq:interface-coupling}
\end{equation}
For a topologically trivial material interfaced with a topological insulator, $\theta_1=0$ and $\theta_2=\pi \bmod 2\pi$, so that $\alpha=\alpha_{\mathrm{fs}}$. For an insulating, magnetically gapped interface, the associated surface Hall conductivity is
$\sigma_H=(\Delta\theta/2\pi)e^2/h$, up to an integer multiple of $e^2/h$. The resulting boundary response couples TE and TM polarizations. Quantized Faraday and Kerr rotations associated with this response have been observed under appropriate magnetic-field conditions~\cite{wu2016quantized}.

Hyperbolic materials~(HMs), in which the principal dielectric permittivities have opposite signs, support electromagnetic modes with large wave vectors and have attracted intense interest for subwavelength imaging, enhanced spontaneous emission, and waveguiding~\cite{caldwell2014sub,poddubny2013hyperbolic,caldwell2015low}. Artificial HMs based on alternating metallic and dielectric layers were among the first systems proposed~\cite{smith2003electromagnetic,jacob2006optical}. More recently, natural hyperbolic materials, most notably hexagonal boron nitride (hBN)~\cite{dai2014tunable} and $\alpha$-MoO$_3$~\cite{zheng2019mid,ma2018plane}, have emerged as low-loss alternatives with clearly resolved phonon polariton resonances in the mid-infrared and terahertz spectral ranges. In polar crystals, these resonances occur within Reststrahlen bands, which are the spectral intervals between transverse optical and longitudinal optical phonon frequencies where a relevant principal permittivity component becomes negative. In anisotropic crystals, different tensor components can have different Reststrahlen bands, and this response enables natural hyperbolicity.

Surface electromagnetic waves at topological insulators~(TIs) interfaces have been studied in several specific configurations, including surface plasmon polaritons in TIs~\cite{di2013observation}, Dyakonov-Tamm waves at interfaces between TIs and structurally chiral materials~\cite{chiadini2016left}, and hybrid TE-TM surface waves at interfaces between TIs and dielectric media~\cite{maimistov2016surface}.

To the best of our knowledge, no study has computed the surface wave dispersion, identified band gaps, or analyzed penetration depth behavior for realistic combinations of HM/TI, such as hBN/Bi$_2$Se$_3$ or $\alpha$-MoO$_3$/Bi$_2$Se$_3$. This gap is significant because the frequency dependent permittivities of real materials introduce qualitatively new behavior, including spectral band gaps, divergences of penetration depth, and propagation windows that depend strongly on the material parameters. These effects are not visible in the parametric analysis of~\cite{lyashko2017surface}.

In addition to natural hyperbolic crystals, we also consider a realistic artificial hyperbolic media based on a dissipative titanium and silicon composite. This case is important because artificial hyperbolic media are generally lossy, so the surface wave propagation constant becomes complex valued. As a result, satisfying the dispersion relation is not by itself sufficient to identify a physical guided mode. One must also verify decay away from the interface, finite penetration depths, and propagation over distances that exceed the modal confinement scale.

In this work, we build on the analytical framework of~\cite{lyashko2017surface} and derive closed form criteria for the existence and localization of surface waves at HM/TI interfaces. The analysis distinguishes the roles of the relevant physical contributions: the permittivity mismatch provides the dominant contribution to the band gap boundaries, the bulk axion coupling produces a small correction of order $\alpha^2$ to both the dispersion and the band edge, and surface Hall responses in magnetic TIs offer a possible mechanism for stronger nonreciprocal effects. We then evaluate these criteria for realistic mid-infrared material systems. Natural phonon polaritonic materials are used to illustrate the lossless band gap structure, while a dissipative titanium and silicon effective medium interfaced with Bi$_2$Se$_3$ is used to show how physical decay conditions, attenuation, and energy flow diagnostics enter in lossy systems.

\section{Model and boundary conditions}
\label{sec:model}

In the bulk of a topological insulator with a spatially uniform axion parameter $\theta$, the macroscopic Maxwell equations retain their conventional form. The topological contribution becomes observable at interfaces, where $\theta$ changes discontinuously and modifies the electromagnetic boundary conditions. In Gaussian units, the constitutive relations including the axion contribution can be written as
\begin{equation}
	\mathbf{D}_a
	=
	\mathbf{E}
	+
	4\pi\mathbf{P}
	-
	\frac{\alpha_{\mathrm{fs}}\theta}{\pi}\mathbf{B},
	\;
	\mathbf{H}_a
	=
	\mathbf{B}
	-
	4\pi\mathbf{M}
	+
	\frac{\alpha_{\mathrm{fs}}\theta}{\pi}\mathbf{E}.
\end{equation}
Here $\alpha_{\mathrm{fs}}=e^2/\hbar c$ is the fine-structure constant. The signs are chosen consistently with the boundary-condition convention used below.

The hyperbolic material is modeled as a uniaxial anisotropic medium with optical axis $\mathbf{l}$. Its electric induction is
\begin{equation}
	\mathbf{D} = \varepsilon_o\mathbf{E}
	+ \left(\varepsilon_e-\varepsilon_o\right)
	\left(\mathbf{l}\cdot\mathbf{E}\right)\mathbf{l},
\end{equation}
where $\varepsilon_o$ and $\varepsilon_e$ are the ordinary and extraordinary permittivities. The medium is hyperbolic when these two principal permittivities have opposite signs. In the convention used here, $\varepsilon_o<0$ and $\varepsilon_e>0$ correspond to a type-I hyperbolic response, whereas $\varepsilon_o>0$ and $\varepsilon_e<0$ correspond to a type-II response. We consider a planar interface at $x=0$, with the hyperbolic material occupying $x<0$ and the topological insulator occupying $x>0$. The surface wave propagates along the $z$ direction and is assumed to be independent of $y$.

\begin{figure}
	\centering
	\includegraphics[width=0.7\linewidth]{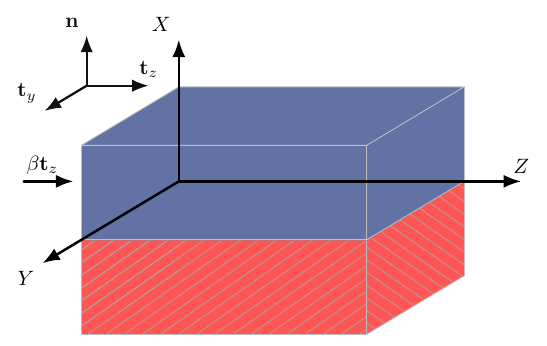}
	\caption{Interface geometry for the hyperbolic material (HM) and topological insulator (TI) structure. The HM occupies $x<0$ and is shown with hatching, while the TI occupies $x>0$. The surface wave propagates along the $z$ direction with propagation constant $\beta$. The unit normal $\mathbf n$ points from the HM into the TI, and $\mathbf t_y$ and $\mathbf t_z$ denote tangential directions along the interface.}
	\label{fig:interface-schematic}
\end{figure}

Figure~\ref{fig:interface-schematic} defines the interface geometry used throughout the analysis. The unit normal $\mathbf n$ is directed from the hyperbolic material toward the topological insulator, and the surface wave propagates along the interface with propagation constant $\beta$ in the $\mathbf t_z$ direction. We consider several orientations of the optical axis $\mathbf l$, since tangential, normal, and longitudinal alignments lead to different decay conditions and therefore to different regimes of surface wave existence.

The axion contribution modifies the electromagnetic boundary conditions at the interface. With superscripts $(1)$ and $(2)$ denoting the hyperbolic material and the topological insulator, respectively, these conditions are
\begin{equation}
	\begin{aligned}
		(\mathbf{D}^{(1)}-\mathbf{D}^{(2)})\cdot\mathbf{n}
		&= -\alpha\,\mathbf{B}^{(1)}\cdot\mathbf{n},
		\\
		(\mathbf{B}^{(1)}-\mathbf{B}^{(2)})\cdot\mathbf{n}
		&= 0,
		\\
		(\mathbf{H}^{(1)}-\mathbf{H}^{(2)})\cdot\mathbf{t}_{z,y}
		&= \alpha\,\mathbf{E}^{(1)}\cdot\mathbf{t}_{z,y},
		\\
		(\mathbf{E}^{(1)}-\mathbf{E}^{(2)})\cdot\mathbf{t}_{z,y}
		&= 0.
	\end{aligned}
	\label{eq:boundary-conditions}
\end{equation}

The interface coupling $\alpha=\alpha_{\mathrm{fs}}\Delta\theta/\pi$ mixes TE and TM polarizations. For the interface considered here, $\Delta\theta=\pi$, and therefore
$\alpha=\alpha_{\mathrm{fs}}\approx1/137$. This polarization mixing is a characteristic feature of topological insulator electrodynamics and is absent at conventional dielectric interfaces.

For a surface wave propagating along $z$ with propagation constant $\beta$, the bulk field equations separate into TE components, $(E_y,H_x,H_z)$, and TM components, $(H_y,E_x,E_z)$. In an isotropic region, the corresponding scalar amplitudes satisfy
\begin{equation}
	\frac{\partial^2 E_y}{\partial z^2} + \frac{\partial^2 E_y}{\partial x^2} + k_0^2\varepsilon E_y = 0,
\end{equation}
for the TE component and
\begin{equation}
	\frac{\partial^2 H_y}{\partial z^2} + \frac{\partial^2 H_y}{\partial x^2} + k_0^2\varepsilon H_y = 0,
\end{equation}
for the TM component. In the hyperbolic medium, the effective permittivity entering these equations depends on the orientation of the optical axis.

\section{Dispersion relations}
\label{sec:dispersion}

We first consider the tangential-axis configuration $\mathbf l=\mathbf t_y$, for which
\begin{equation}
	D_x=\varepsilon_o E_x,\quad
	D_y=\varepsilon_e E_y,\quad
	D_z=\varepsilon_o E_z .
\end{equation}
Surface wave solutions decaying away from the interface are written as
\begin{equation}
	\begin{aligned}
			E_y^{(1)} &= A\exp(p_1 x+i\beta z),
			&
			p_1^2 &= \beta^2-k_0^2\varepsilon_e > 0,
			\\
			H_y^{(1)} &= B\exp(p_2 x+i\beta z),
			&
			p_2^2 &= \beta^2-k_0^2\varepsilon_o > 0,
			\\
			E_y^{(2)} &= C\exp(-qx+i\beta z),
			&
			q^2 &= \beta^2-k_0^2\varepsilon_2 > 0,
			\\
			H_y^{(2)} &= F\exp(-qx+i\beta z).
		\end{aligned}
	\label{eq:surface-wave-ansatz}
\end{equation}
Here superscripts $(1)$ and $(2)$ denote the hyperbolic material and the topological insulator, respectively. Applying the boundary conditions gives
\begin{equation}
	(q + p_1)\left(\frac{q}{\varepsilon_2} + \frac{p_2}{\varepsilon_o}\right) + \alpha^2\frac{p_2 q}{\varepsilon_o\varepsilon_2} = 0 .
	\label{eq:disp1}
\end{equation}
A localized solution requires $\varepsilon_o<0$, so the relevant hyperbolic regime is the type-I case. In this regime Eq.~(\ref{eq:disp1}) can be written as
\begin{equation}
	(q + p_1)\left(\frac{q}{\varepsilon_2} - \frac{p_2}{|\varepsilon_o|}\right) - \alpha^2\frac{p_2 q}{|\varepsilon_o|\varepsilon_2} = 0 .
	\label{eq:disp2}
\end{equation}

For the normal-axis configuration $\mathbf l=\mathbf n$, the hyperbolic medium tensor components entering the field equations become
\begin{equation}
	D_x=\varepsilon_e E_x,\quad
	D_y=\varepsilon_o E_y,\quad
	D_z=\varepsilon_o E_z .
\end{equation}
For fields proportional to $\exp(p_jx+i\beta z)$, the TE-like and
TM-like decay constants in the hyperbolic medium are
\begin{equation}
	\begin{aligned}
		p_1^2 &= \beta^2-k_0^2\varepsilon_o,\\
		p_2^2 &=
		\frac{\varepsilon_o}{\varepsilon_e}
		\left(\beta^2-k_0^2\varepsilon_e\right).
	\end{aligned}
	\label{eq:normal-axis-decay}
\end{equation}
The resulting dispersion relation retains the same algebraic form as
Eq.~(\ref{eq:disp1}), but with the decay constants appropriate to the
normal-axis geometry. In the type-I regime,
$\varepsilon_o<0$ and $\varepsilon_e>0$, the condition $p_1^2>0$ is
automatically satisfied for real $\beta$, whereas $p_2^2>0$ requires
\begin{equation}
	\beta^2<k_0^2\varepsilon_e.
\end{equation}
Together with the TI-side decay condition
$q^2=\beta^2-k_0^2\varepsilon_2>0$, this gives the localized interval
\begin{equation}
	\varepsilon_2<n^2<\varepsilon_e,
\end{equation}
which can exist only when $\varepsilon_e>\varepsilon_2$.

If the optical axis is instead parallel to the propagation direction, $\mathbf l=\mathbf t_z$, one obtains $p_2^2=(\varepsilon_e/\varepsilon_o)p_1^2$. In a hyperbolic medium, $\varepsilon_e/\varepsilon_o<0$, so this choice gives an oscillatory rather than evanescent TM-like field in the hyperbolic half-space. The propagation-axis configuration therefore does not support a localized surface wave.

We introduce the lossless effective index $n=\beta/k_0$, which is real in the absence of dissipation. In lossy media, the corresponding complex quantity will be denoted by $n_{\mathrm{eff}}$. For the tangential-axis configuration in the type-I regime,
$\varepsilon_o<0$, the decay constants take the form
\begin{equation}
	\begin{aligned}
		q &= k_0\sqrt{n^2-\varepsilon_2},\\
		p_1 &= k_0\sqrt{n^2-\varepsilon_e},\\
		p_2 &= k_0\sqrt{n^2+|\varepsilon_o|}.
	\end{aligned}
	\label{eq:decay-constants-index}
\end{equation}
For the normal-axis configuration, the corresponding lossless decay
constants are instead
\begin{equation}
	\begin{aligned}
		q &= k_0\sqrt{n^2-\varepsilon_2},\\
		p_1 &= k_0\sqrt{n^2-\varepsilon_o},\\
		p_2 &= k_0\sqrt{
			\frac{\varepsilon_o}{\varepsilon_e}
			\left(n^2-\varepsilon_e\right)}.
	\end{aligned}
	\label{eq:decay-constants-normal-index}
\end{equation}
The positive real square roots are selected in the lossless localized
interval $\varepsilon_2<n^2<\varepsilon_e$. Substitution into Eq.~(\ref{eq:disp2}) gives
\begin{widetext}
\begin{equation}
	\left(\sqrt{n^2-\varepsilon_2}+\sqrt{n^2-\varepsilon_e}\right)
	\left(
	\frac{\sqrt{n^2-\varepsilon_2}}{\varepsilon_2}
	-
	\frac{\sqrt{n^2+|\varepsilon_o|}}{|\varepsilon_o|}
	\right)
	=
	\frac{\alpha^2}{|\varepsilon_o|\varepsilon_2}
	\sqrt{(n^2-\varepsilon_2)(n^2+|\varepsilon_o|)} ,
	\label{eq:disp-norm}
\end{equation}
\end{widetext}
with the tangential-axis decay conditions $n^2>\varepsilon_2$ and $n^2>\varepsilon_e$.

\section{Band gap analysis}
\label{sec:bandgap}

\subsection{Lossless case}
\label{sec:bandgap-lossless}

We first consider the ideal lossless case, in which permittivities $\varepsilon_o$, $\varepsilon_e$, and $\varepsilon_2$ are real. In this limit the propagation constant $\beta$ and the effective index $n$ are also real, and the existence of a localized surface wave is determined by two requirements: exponential decay away from the interface and satisfaction of the dispersion relation.

For the tangential optical axis configuration, $\mathbf l=\mathbf t_y$, decay into the TI requires
\begin{equation}
	q^2>0,
	\quad \text{or equivalently} \quad
	n^2>\varepsilon_2 .
\end{equation}
The TE component in the hyperbolic material must also decay, which gives
\begin{equation}
	p_1^2>0,
	\quad \text{or equivalently} \quad
	n^2>\varepsilon_e .
\end{equation}
The TM decay condition in the hyperbolic material is
\begin{equation}
	p_2^2>0,
	\quad \text{or equivalently} \quad
	n^2>-|\varepsilon_o| .
\end{equation}
In the type-I hyperbolic regime, $\varepsilon_o<0$, this last condition is automatically satisfied for physical modes with $n^2>0$. The nontrivial decay requirements therefore impose the lower bound $n^2>\max\{\varepsilon_2,\varepsilon_e\}$. The surface wave exists only when this admissible interval is compatible with the dispersion equation~(\ref{eq:disp2}). When the material permittivities depend on frequency, both the lower bound and the solvability of the dispersion relation become frequency dependent, so the spectrum separates into allowed surface wave intervals and forbidden intervals.

It is useful to isolate the sign condition that controls solvability. Dividing Eq.~(\ref{eq:disp2}) by $p_2q/(|\varepsilon_o|\varepsilon_2)$ gives
\begin{equation}
	(q + p_1)\left(\frac{|\varepsilon_o|}{p_2} - \frac{\varepsilon_2}{q}\right) = \alpha^2.
	\label{eq:ratio-form}
\end{equation}

Define the effective ratio
\begin{equation}
	R(n^2, \omega) \equiv \frac{|\varepsilon_o(\omega)|}{p_2(n^2, \omega)} - \frac{\varepsilon_2(\omega)}{q(n^2, \omega)},
	\label{eq:R-def}
\end{equation}
where $p_2 = k_0\sqrt{n^2 + |\varepsilon_o|}$ and $q = k_0\sqrt{n^2 - \varepsilon_2}$. The dispersion equation~(\ref{eq:ratio-form}) requires
\begin{equation}
	R(n^2, \omega) = \frac{\alpha^2}{q + p_1} > 0.
	\label{eq:R-positive}
\end{equation}

Since $\alpha^2>0$ and $q+p_1>0$, positivity of
$R(n^2,\omega)$ is a necessary, but not by itself sufficient,
condition for the existence of a surface wave. To state the exact
dispersion condition, define
\begin{equation}
	F(n^2,\omega)
	\equiv
	(q+p_1)R(n^2,\omega).
	\label{eq:F-def}
\end{equation}
Equation~(\ref{eq:ratio-form}) is then equivalent to
\begin{equation}
	F(n^2,\omega)=\alpha^2.
	\label{eq:F-dispersion}
\end{equation}

At fixed frequency, the admissible effective indices satisfy
\begin{equation}
	n^2>n_{\min}^2,
	\qquad
	n_{\min}^2=
	\max\{\varepsilon_2(\omega),\varepsilon_e(\omega)\}.
\end{equation}
A localized surface wave exists only if Eq.~(\ref{eq:F-dispersion})
has a solution in this interval. Therefore, a sufficient condition
for the absence of a surface wave is
\begin{equation}
	\sup_{n^2>n_{\min}^2}F(n^2,\omega)<\alpha^2,
\end{equation}
while the boundary of the existence region is reached when the
supremum equals $\alpha^2$.

For the material regime considered below,
$\varepsilon_2\geq\varepsilon_e$, the relevant upper end of the
surface wave branch occurs in the large-effective-index limit.
Using
\begin{equation}
	q\sim k_0n,\qquad
	p_1\sim k_0n,\qquad
	p_2\sim k_0n
\end{equation}
as $n^2\to\infty$, we obtain
\begin{equation}
	\lim_{n^2\to\infty}F(n^2,\omega)
	=
	2\left[
	|\varepsilon_o(\omega)|-\varepsilon_2(\omega)
	\right].
\end{equation}
Consequently, the large-index band edge satisfies
\begin{equation}
	2\left[
	|\varepsilon_o(\omega_*)|-\varepsilon_2(\omega_*)
	\right]
	=
	\alpha^2,
\end{equation}
or equivalently
\begin{equation}
	|\varepsilon_o(\omega_*)|
	=
	\varepsilon_2(\omega_*)+\frac{\alpha^2}{2}.
	\label{eq:band-edge-corrected}
\end{equation}
Thus, the dielectric contrast provides the dominant contribution to
the band edge, while the topological coupling shifts it by the small
quantity $\alpha^2/2$. The condition
$|\varepsilon_o|=\varepsilon_2$ is recovered only in the
non-topological limit $\alpha=0$.

Accordingly, for the surface wave branch and parameter regime considered here, the lossless band gap is described by
\begin{equation}
	|\varepsilon_o(\omega)|
	<
	\varepsilon_2(\omega)+\frac{\alpha^2}{2},
	\label{eq:bandgap}
\end{equation}
for which no localized surface wave solution exists. The equality in Eq.~(\ref{eq:band-edge-corrected}) defines the band edge, where the effective index diverges in the local lossless model.

For a hyperbolic material with ordinary permittivity given by the Lorentz model
\begin{equation}
	\varepsilon_o(\omega) = \varepsilon_{\infty,o}\frac{\omega_{\mathrm{LO},o}^2 - \omega^2}{\omega_{\mathrm{TO},o}^2 - \omega^2},
\end{equation}
where $\omega_{\mathrm{TO},o}$ and $\omega_{\mathrm{LO},o}$ are the ordinary transverse- and longitudinal-optical phonon frequencies. The hyperbolic regime ($\varepsilon_o < 0$) occupies the Reststrahlen band $\omega_{\mathrm{TO},o} < \omega < \omega_{\mathrm{LO},o}$. Within this band, we have
\begin{equation}
	|\varepsilon_o(\omega)| = \varepsilon_{\infty,o}\frac{\omega_{\mathrm{LO},o}^2 - \omega^2}{\omega^2 - \omega_{\mathrm{TO},o}^2}.
\end{equation}

For compactness, define
\begin{equation}
	\varepsilon_{2,\alpha}
	\equiv
	\varepsilon_2+\frac{\alpha^2}{2}.
	\label{eq:eps2-alpha}
\end{equation}

Setting
$|\varepsilon_o(\omega_*)|=\varepsilon_{2,\alpha}$
and solving for $\omega_*$ within the Reststrahlen band gives
\begin{equation}
	\omega_*^2
	=
	\frac{
		\varepsilon_{\infty,o}\omega_{\mathrm{LO},o}^2
		+
		\varepsilon_{2,\alpha}\omega_{\mathrm{TO},o}^2
	}{
		\varepsilon_{\infty,o}
		+
		\varepsilon_{2,\alpha}
	}.
	\label{eq:omega-star}
\end{equation}

This frequency always lies within the Reststrahlen band, $\omega_{\mathrm{TO}}<\omega_*<\omega_{\mathrm{LO}}$, and separates the band into an allowed surface wave interval and a forbidden interval. For $\omega_{\mathrm{TO}}<\omega<\omega_*$, the magnitude of the ordinary permittivity satisfies $|\varepsilon_o|>\varepsilon_{2,\alpha}$, so localized surface waves can exist; this is the region near the TO frequency, where $|\varepsilon_o|$ is large and diverges in the lossless Lorentz model. For $\omega_*<\omega<\omega_{\mathrm{LO}}$, the inequality is reversed, $|\varepsilon_o|<\varepsilon_{2,\alpha}$, and the surface wave is forbidden. This upper part of the Reststrahlen band therefore forms the surface wave band gap, with $|\varepsilon_o|\to 0$ as $\omega\to\omega_{\mathrm{LO}}$.

The surface wave (SW) bandwidth fraction, denoted by $f_{\mathrm{SW}}$, is
\begin{equation}
	f_{\mathrm{SW}} = \frac{\omega_* - \omega_{\mathrm{TO}}}{\omega_{\mathrm{LO}} - \omega_{\mathrm{TO}}}.
	\label{eq:f-SW}
\end{equation}

When $\varepsilon_2 \gg \varepsilon_{\infty,o}$, as in Bi$_2$Se$_3$ with $\varepsilon_2 \approx 41$ compared with hBN with $\varepsilon_{\infty,o} \approx 4.87$, the frequency $\omega_*$ is shifted toward $\omega_{\mathrm{TO}}$ and the surface wave window becomes narrow. This narrowing is a direct consequence of the large dielectric response of the TI relative to the background permittivity of the hyperbolic material.

Physically, the surface wave requires the ordinary permittivity of the hyperbolic material to be sufficiently negative to overcome the positive permittivity of the TI. Since the Reststrahlen band is the interval in which the phonon resonance drives $\varepsilon_o(\omega)$ through negative values, it provides the natural spectral window in which such a surface wave can exist. Near the TO frequency, $|\varepsilon_o| \to \infty$, so this requirement is easily satisfied. Near the LO frequency, $|\varepsilon_o| \to 0$, so it is not satisfied. The crossover at $\omega_*$ separates the two regimes.

Equation~(\ref{eq:band-edge-corrected}) shows that the topological coupling shifts the band edge through the term $\alpha^2/2$. This correction is extremely small for conventional TI parameters, so the band edge is governed overwhelmingly by the dielectric contrast. Nevertheless, the correction is finite and should be retained when stating the analytical existence condition.

Within the allowed frequency interval, $\alpha$ also modifies the effective index and penetration depths. Both the shift of the band edge and the modification of the dispersion scale as $\alpha^2$ and are therefore small for conventional TI parameters. Physically, the axion coupling mixes the TE and TM field components at the interface, producing a weak correction to the propagation constant and to the frequency interval over which the localized branch exists.

Thus, in the lossless axion boundary model, the surface wave existence region is controlled primarily by the permittivity mismatch, with an additional band-edge shift of order $\alpha^2$ from the topological coupling.

\subsection{Lossy case}
\label{sec:bandgap-lossy}

In dissipative media, the permittivities are complex, and the lossless existence criterion must be replaced by a selection criterion for complex solutions of the dispersion relation. The dispersion relation retains the same algebraic form, but the effective index takes complex values. A mathematical solution of the dispersion relation represents a physical surface wave only if the fields decay away from the interface and attenuate along the direction of propagation. Thus, at each frequency, the admissible solution must satisfy
\begin{equation}
	\operatorname{Re}q>0,\quad
	\operatorname{Re}p_1>0,\quad
	\operatorname{Re}p_2>0,\quad
	\operatorname{Im}\beta>0,
	\label{eq:complex-branch-selection-main}
\end{equation}
where first three inequalities enforce spatial localization in the transverse direction, while the last one selects waves attenuating in the positive $z$ direction.

The complex square roots entering $q$, $p_1$, and $p_2$ introduce a branch selection issue. We denote the complex arguments of these square roots by $z_q$, $z_{p_1}$, and $z_{p_2}$. Here the symbol $z$ is used only as a conventional notation for a complex variable and should not be confused with the spatial coordinate. For example, the TI-side decay constant is determined by the square root argument
\begin{equation}
	z_q=n_{\mathrm{eff}}^2-\varepsilon_2.
\end{equation}
The hyperbolic-medium square-root arguments depend on the optical-axis
orientation. For the tangential-axis configuration they are
\begin{equation}
	z_{p_1}=n_{\mathrm{eff}}^2-\varepsilon_e,
	\qquad
	z_{p_2}=n_{\mathrm{eff}}^2-\varepsilon_o .
\end{equation}
For the normal-axis configuration they are instead
\begin{equation}
	z_{p_1}=n_{\mathrm{eff}}^2-\varepsilon_o,
	\qquad
	z_{p_2}=
	\frac{\varepsilon_o}{\varepsilon_e}
	\left(n_{\mathrm{eff}}^2-\varepsilon_e\right).
\end{equation}
In each configuration, the square-root branches are selected so that
$\operatorname{Re}p_1>0$ and $\operatorname{Re}p_2>0$.
As the frequency changes, these complex arguments trace trajectories in the complex plane and may approach or cross a branch cut of the square root. When this occurs, the square root branch must be chosen so that the decay and attenuation conditions in Eq.~(\ref{eq:complex-branch-selection-main}) remain satisfied. Apparent jumps or gaps in a tracked effective index curve can therefore arise from the branch structure of the complex square roots in the lossy dispersion problem.

\section{Penetration depth and divergence conditions}
\label{sec:penetration}

The surface wave has three characteristic transverse penetration depths, each associated with decay constant
\begin{equation}
	\delta_{\mathrm{TI}}=\frac{1}{\operatorname{Re}q},
	\quad
	\delta_{1}=\frac{1}{\operatorname{Re}p_1},
	\quad
	\delta_{2}=\frac{1}{\operatorname{Re}p_2}.
	\label{eq:penetration-depths}
\end{equation}
Here $\delta_{\mathrm{TI}}$ is the penetration depth into the topological insulator, while $\delta_1$ and $\delta_2$ are the two hyperbolic medium penetration depths associated with the TE-like and TM-like decay constants $p_1$ and $p_2$, respectively. In the lossless limit, $q$, $p_1$, and $p_2$ are real, so Eq.~(\ref{eq:penetration-depths}) becomes
\begin{equation}
	\delta_{\mathrm{TI}}=\frac{1}{q},
	\quad
	\delta_1=\frac{1}{p_1},
	\quad
	\delta_2=\frac{1}{p_2}.
	\label{eq:penetration-depths-lossless}
\end{equation}

The penetration depth $\delta_2$ remains finite throughout the type-I hyperbolic regime. Indeed, in this regime $p_2^2=k_0^2(n^2+|\varepsilon_o|)$, which cannot vanish for real physical $n^2>0$. Thus, possible delocalization can occur only when either $q$ or $p_1$ approaches zero.

For lossy modes, the propagation constant also defines a longitudinal intensity attenuation length,
\begin{equation}
	L_{\mathrm{att}}=\frac{1}{2\operatorname{Im}\beta}.
	\label{eq:attenuation-length-main}
\end{equation}
Together, $L_{\mathrm{att}}$, $\delta_{\mathrm{TI}}$, $\delta_1$, and $\delta_2$ characterize whether a complex dispersion root corresponds to a localized guided mode with appreciable propagation distance.

\subsection{Divergence conditions}
\label{sec:divergence}

We first examine possible loss of confinement on the TI side. In the lossless case $\delta_{\mathrm{TI}}=1/q$, so TI-side delocalization would require $q\to0$, or equivalently $n^2\to\varepsilon_2$. Substituting $q=0$ into Eq.~(\ref{eq:disp2}) gives $-p_1p_2/|\varepsilon_o|=0$, which cannot be satisfied on a localized type-I branch, where $p_2>0$ and $p_1>0$. Thus, at a regular point of the lossless dispersion curve, the TI-side penetration depth does not diverge by itself. In the non-topological limit, $\alpha=0$, the dispersion relation factors as
\begin{equation}
	(q+p_1)\left(\frac{q}{\varepsilon_2}-\frac{p_2}{|\varepsilon_o|}\right)=0 .
	\label{eq:alpha-zero}
\end{equation}
The first factor gives no localized solution because $q,p_1>0$. The second factor determines the physical branch and shows that the limit $q\to0$ can be approached only asymptotically, for example near a material pole where $|\varepsilon_o|$ becomes large. This is distinct from the band gap boundary $|\varepsilon_o|=\varepsilon_2$, which is reached through the large-$n$ limit discussed above.

We next consider delocalization of the TE-like component into the hyperbolic medium. This occurs when $p_1\to0$, or $n^2\to\varepsilon_e$. Setting $p_1=0$ in Eq.~(\ref{eq:disp2}) yields
\begin{equation}
	q\left(\frac{q}{\varepsilon_2}-\frac{p_2}{|\varepsilon_o|}\right)
	=
	\alpha^2\frac{p_2q}{|\varepsilon_o|\varepsilon_2}.
	\label{eq:div-TE-1}
\end{equation}
Since $q\neq0$ at this point provided $\varepsilon_e>\varepsilon_2$, this reduces to
\begin{equation}
	\frac{q}{\varepsilon_2}
	=
	\frac{p_2}{|\varepsilon_o|}
	\left(1+\frac{\alpha^2}{\varepsilon_2}\right).
\end{equation}
Using $n^2=\varepsilon_e$ gives $q=k_0\sqrt{\varepsilon_e-\varepsilon_2}$ and $p_2=k_0\sqrt{\varepsilon_e+|\varepsilon_o|}$. The TE-delocalization condition is therefore
\begin{equation}
	\frac{\sqrt{\varepsilon_e-\varepsilon_2}}{\varepsilon_2}
	=
	\frac{\sqrt{\varepsilon_e+|\varepsilon_o|}}{|\varepsilon_o|}
	\left(1+\frac{\alpha^2}{\varepsilon_2}\right).
	\label{eq:div-freq}
\end{equation}
When this equation is satisfied, the TE-like field ceases to decay into the hyperbolic medium and the corresponding penetration depth $\delta_1$ diverges. A real delocalization frequency $\omega_{\mathrm{div}}$ can occur only in spectral ranges where $\varepsilon_e>\varepsilon_2$ and Eq.~(\ref{eq:div-freq}) is satisfied.

For Lorentz oscillator permittivities, Eq.~(\ref{eq:div-freq}) gives an implicit algebraic condition for the TE-delocalization frequency $\omega_{\mathrm{div}}$. In the non-topological limit, $\alpha\to0$, this condition reduces to
\begin{equation}
	\frac{|\varepsilon_o(\omega)|}{\varepsilon_2}
	=
	\sqrt{
		\frac{\varepsilon_e(\omega)+|\varepsilon_o(\omega)|}
		{\varepsilon_e(\omega)-\varepsilon_2}
	}.
	\label{eq:div-simplified}
\end{equation}
This equation can be solved numerically for each material pair. The finite-$\alpha$ correction enters through the factor $(1+\alpha^2/\varepsilon_2)$ in Eq.~(\ref{eq:div-freq}) and is therefore small for conventional TI parameters, since $\alpha^2\approx5.3\times10^{-5}$.

The divergence of penetration depth marks a transition from surface wave to leaky wave. At $\omega_{\mathrm{div}}$, the TE component ceases to be confined and radiates into the bulk of the hyperbolic medium. Above $\omega_{\mathrm{div}}$ (or below, depending on the dispersion), the surface wave persists but with only TM confinement in the HM side.

In the type-I lossless regime, the TM-like decay constant in the hyperbolic medium is always larger than both other transverse decay constants, since
\begin{equation}
	\begin{aligned}
		p_2^2-q^2 = k_0^2\left(|\varepsilon_o|+\varepsilon_2\right) &> 0,\\
		p_2^2-p_1^2 = k_0^2\left(|\varepsilon_o|+\varepsilon_e\right) &> 0.
	\end{aligned}
\end{equation}
Thus $\delta_2$ is always the smallest penetration depth. The relative ordering of $\delta_{\mathrm{TI}}$ and $\delta_1$ is not universal, because
\begin{equation}
	p_1^2-q^2 = k_0^2\left(\varepsilon_2-\varepsilon_e\right).
\end{equation}
Therefore, the penetration-depth hierarchy is
\begin{equation}
	\begin{cases}
		\delta_2 < \delta_1 < \delta_{\mathrm{TI}},
		& \varepsilon_2>\varepsilon_e,\\[3pt]
		\delta_2 < \delta_{\mathrm{TI}} < \delta_1,
		& \varepsilon_e>\varepsilon_2.
	\end{cases}
	\label{eq:hierarchy}
\end{equation}
For the hBN/Bi$_2$Se$_3$ parameters used below, $\varepsilon_2>\varepsilon_e$, so the expected ordering is $\delta_2<\delta_1<\delta_{\mathrm{TI}}$.

The total surface wave thickness (full width at half maximum of energy density) is dominated by the largest penetration depth
\begin{equation}
	\delta_{\mathrm{total}} \approx \delta_1 + \delta_{\mathrm{TI}}.
	\label{eq:delta-total}
\end{equation}
This quantity provides an estimate of the modal thickness probed in near-field measurements.

\section{Numerical results}
\label{sec:numerical}

The natural material calculations use hBN, $\alpha$-MoO$_3$, and SiC as representative phonon polaritonic hyperbolic media, with Bi$_2$Se$_3$ as the topological insulator. For hBN, we use a Lorentz model with ordinary high frequency permittivity $\varepsilon_{\infty,o}=4.87$ and extraordinary high frequency permittivity $\varepsilon_{\infty,e}=2.95$, based on the first principles dielectric constants reported by Cai~et~al.~\cite{cai2007infrared}. The TO and LO phonon frequencies are taken from the optical constant fit for hBN by Caldwell~et~al.~\cite{caldwell2015low}. We use the upper Reststrahlen band, 1360--1614~cm$^{-1}$, where $\varepsilon_o<0$ and $\varepsilon_e>0$. This interval corresponds to the type-I hyperbolic regime relevant to the localized surface wave solutions derived above. The lower Reststrahlen band of hBN, 760--825~cm$^{-1}$, has type-II character and is therefore not used for the localized branches considered here.

For $\alpha$-MoO$_3$, which is a biaxial crystal, we use an effective uniaxial reduction of the dielectric tensor. The negative component $\varepsilon_o$ is identified with the $[100]$ principal component, using $\varepsilon_{\infty,o}=4.0$,
$\omega_{\mathrm{TO},o}=820~\mathrm{cm}^{-1}$, and $\omega_{\mathrm{LO},o}=972~\mathrm{cm}^{-1}$. The positive component $\varepsilon_e$ is identified with the $[010]$ principal component, using $\varepsilon_{\infty,e}=2.4$,
$\omega_{\mathrm{TO},e}=958~\mathrm{cm}^{-1}$, and $\omega_{\mathrm{LO},e}=1004~\mathrm{cm}^{-1}$. This choice gives $\varepsilon_o<0$ and $\varepsilon_e>0$ throughout the predicted surface-wave window. Because $\alpha$-MoO$_3$ is biaxial, this mapping is used as an effective uniaxial approximation for the band-edge analysis. For Bi$_2$Se$_3$, we use a representative lossless dielectric constant $\varepsilon_2=41$ in the mid-infrared, corresponding to $n\simeq 6.4$, based on the optical constant measurements reported by Nandi~et~al.~\cite{nandi2023high}. Related near-field measurements in the mid-infrared by Menabde~et~al.~support treating bismuth based topological insulators as high-index and low-loss dielectric substrates in this spectral range~\cite{menabde2024high}. SiC is modeled with $\omega_{\mathrm{TO}}=797$~cm$^{-1}$ and $\omega_{\mathrm{LO}}=973$~cm$^{-1}$ from Caldwell~et~al.~\cite{caldwell2015low}, and with $\varepsilon_\infty=6.6$ from Tiwald~et~al.~\cite{tiwald1999carrier}.

In the lossy effective medium approximation for a subwavelength layered structure, the ordinary and extraordinary permittivities are
\begin{equation}
	\varepsilon_o
	=
	f_m\varepsilon_m+(1-f_m)\varepsilon_d,
	\;
	\varepsilon_e
	=
	\frac{\varepsilon_d\varepsilon_m}
	{f_m\varepsilon_d+(1-f_m)\varepsilon_m},
	\label{eq:emt-lossy}
\end{equation}
where $f_m$ is the metal filling fraction, $\varepsilon_m$ is the Ti permittivity, and $\varepsilon_d$ is the Si permittivity. In the numerical calculations, Ti optical constants were taken from the Drude-Lorentz parameterization of Raki\'c~et~al.~\cite{rakic1998optical}, the Si dielectric function was taken from the temperature-dependent optical constants of float-zone crystalline silicon at 20~$^\circ$C reported by Franta~et~al.~\cite{franta2017temperature}, and the Bi$_2$Se$_3$ optical response was taken from broadband optical constants reported by Ermolaev~et~al.~\cite{ermolaev2023broadband}. The representative calculations used $f_m=0.4$.

\subsection{Lossless case, natural hyperbolic materials}

The band gap condition in Eq.~(\ref{eq:bandgap}) restricts localized surface wave propagation to the interval $\omega_{\mathrm{TO}}\leq \omega < \omega_*$, where $\omega_*$ is defined by Eq.~(\ref{eq:omega-star}). The corresponding band edges and propagation windows for the two material systems considered here are summarized in Table~\ref{tab:bandgap}.

\begin{table*}
	\caption{Band gap parameters for surface waves at HM/TI interfaces.}
	\label{tab:bandgap}
	\begin{ruledtabular}
		\begin{tabular*}{\textwidth}{@{\extracolsep{\fill}}lccccc}
			System & Reststrahlen band & $\omega_*$ (cm$^{-1}$) & Surface wave window & Band gap & $f_{\mathrm{SW}}$ \\
			\hline
			hBN / Bi$_2$Se$_3$ & 1360--1614~cm$^{-1}$ & 1389 & 1360--1389 & 1389--1614 & 11.5\% \\
			$\alpha$-MoO$_3$ / Bi$_2$Se$_3$ & 820--972~cm$^{-1}$ & 834 & 820--834 & 834--972 & 9.6\% \\
		\end{tabular*}
	\end{ruledtabular}
\end{table*}

\begin{figure}
	\centering
	\includegraphics[width=1.0\linewidth]{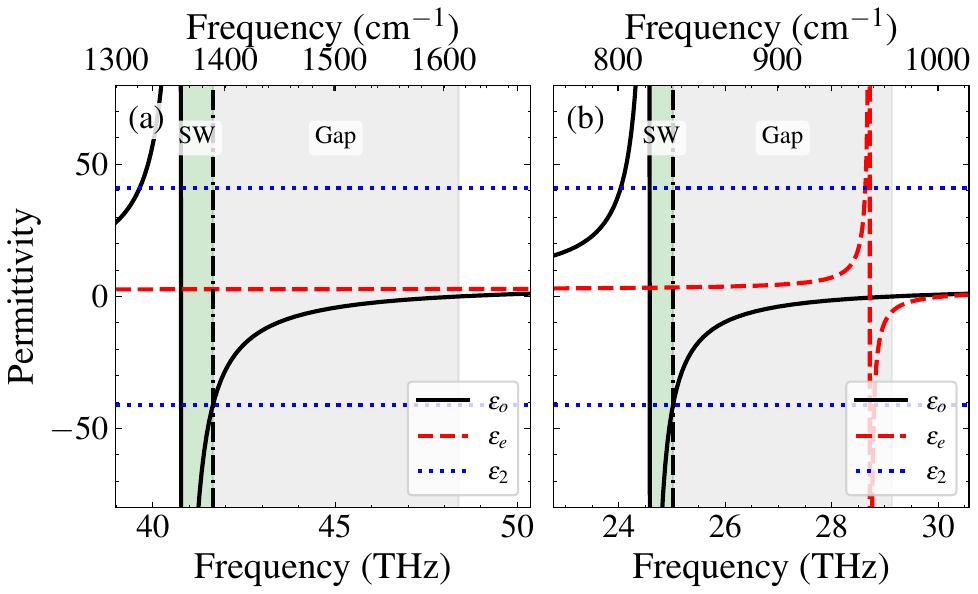}
	\caption{Frequency-dependent permittivities and surface wave (SW) band gap boundaries for natural Reststrahlen-band hyperbolic materials: (a) hBN/Bi$_2$Se$_3$ and (b) $\alpha$-MoO$_3$/Bi$_2$Se$_3$. The curves show the permittivity components denoted by $\varepsilon_o$ and $\varepsilon_e$ in the effective uniaxial model, together with the topological insulator permittivity $\varepsilon_2$. For $\alpha$-MoO$_3$, $\varepsilon_o$ and $\varepsilon_e$ correspond to the selected $[100]$ and $[010]$ principal components, respectively. The mirrored horizontal line at $-\varepsilon_{2,\alpha}$ is included to show the finite-coupling band-edge condition $|\varepsilon_o(\omega)|=\varepsilon_{2,\alpha}$. The green shaded regions indicate the allowed SW windows, while the gray shaded regions indicate the band gaps where $|\varepsilon_o(\omega)|<\varepsilon_{2,\alpha}$. The vertical dash-dotted line marks $\omega_*$, the boundary between the SW window and the band gap.}
	\label{fig:permittivities}
\end{figure}

The resulting surface wave windows are narrow for both systems. This follows from the large dielectric constant of Bi$_2$Se$_3$, $\varepsilon_2 \approx 41$, which shifts $\omega_*$ close to the transverse optical phonon frequency. Figure~\ref{fig:permittivities} shows the frequency dependent permittivities and the corresponding band gap boundaries, while Table~\ref{tab:bandgap} lists the allowed and forbidden intervals.

\begin{figure}
	\centering
	\includegraphics[width=1.0\linewidth]{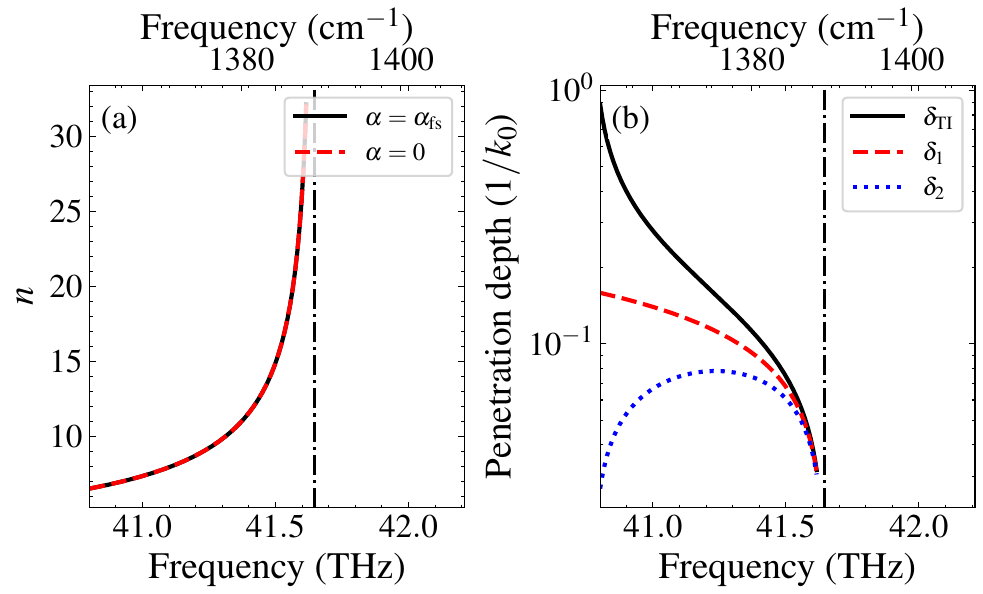}
	\caption{Surface wave dispersion and confinement at the hBN/Bi$_2$Se$_3$ interface within the allowed spectral window. (a) Effective index $n=\beta/k_0$ computed with the topological coupling $\alpha=\alpha_{\mathrm{fs}}$ and in the non-topological limit $\alpha=0$. (b) Penetration depths $\delta_{\mathrm{TI}}$, $\delta_1$, and $\delta_2$, shown in units of $1/k_0$. The vertical dash-dotted line marks $\omega_*$, the upper boundary of the allowed surface wave window.}
	\label{fig:dispersion}
\end{figure}

Figure~\ref{fig:dispersion}(a) shows the effective index $n(\omega)$ for the hBN/Bi$_2$Se$_3$ interface within the surface wave window, 1360--1389~cm$^{-1}$. Near the transverse optical phonon frequency, $|\varepsilon_o|$ becomes large, the TI-side decay constant approaches zero, and the effective index approaches $n\simeq\sqrt{\varepsilon_2}\approx 6.4$. As $\omega$ approaches the finite-coupling band edge $\omega_*$ from below, the condition $|\varepsilon_o|\to\varepsilon_{2,\alpha}$ is reached and the effective index increases rapidly, indicating stronger in-plane confinement. The curves computed with the bulk topological coupling, $\alpha=\alpha_{\mathrm{fs}}$, and in the non-topological limit, $\alpha=0$, are visually indistinguishable on this scale. This confirms that the correction proportional to $\alpha^2\approx5.3\times10^{-5}$ has a negligible effect on the dispersion for the material parameters considered here.

Figure~\ref{fig:dispersion}(b) shows the three penetration depths for hBN/Bi$_2$Se$_3$. Since $\varepsilon_2>\varepsilon_e$ for this material pair, the numerical hierarchy is $\delta_2 < \delta_1 < \delta_{\mathrm{TI}}$, in agreement with Eq.~(\ref{eq:hierarchy}). The largest penetration depth is on the TI side. Near the lower TO edge, $q$ is small and $\delta_{\mathrm{TI}}$ is comparatively large. As $\omega$ approaches $\omega_*$ from below, $n$ increases and the TI-side and TE-like penetration depths decrease, so the mode becomes more tightly confined near the band gap boundary.

On the scale of Fig.~\ref{fig:dispersion}, the curves for $\alpha=\alpha_{\mathrm{fs}}$ and $\alpha=0$ overlap. The corresponding change in both $n$ and the penetration depths is below the plotting resolution for the material parameters used here, consistent with the $\alpha^2$ scaling in Eq.~(\ref{eq:disp-norm}).

The narrow surface wave windows found for bulk Bi$_2$Se$_3$ (11.5\% for hBN, 9.6\% for $\alpha$-MoO$_3$) are a consequence of the large $\varepsilon_2 = 41$ rather than a fundamental limitation. A thin TI film on a low-permittivity substrate reduces the effective $\varepsilon_2$ experienced by the evanescent surface wave field, dramatically expanding the allowed bandwidth.

For a TI film of thickness $d$ smaller than the transverse length scale over which the evanescent field samples the TI side, the field overlaps both the TI film and the dielectric substrate. We model this effect by a simple parallel effective medium estimate
\begin{equation}
	\varepsilon_{2,\mathrm{eff}} =
	\frac{d}{D}\varepsilon_{\mathrm{TI}}
	+
	\left(1-\frac{d}{D}\right)\varepsilon_{\mathrm{sub}},
	\label{eq:eps2eff}
\end{equation}
where $D$ is an effective evanescent sampling length. In the numerical estimates below we use $D=100~\mathrm{nm}$, representative of the near-field penetration scale in the mid-infrared configurations considered here. This estimate is intended to capture the trend that thinner TI films on low-permittivity substrates reduce the effective dielectric constant seen by the surface wave; a full multilayer calculation would be required for quantitative device design.

The band gap boundary is then obtained by replacing $\varepsilon_2$ with $\varepsilon_{2,\mathrm{eff}}$ while retaining the finite topological correction $\alpha^2/2$, giving us
\begin{equation}
	\omega_*^2
	=
	\frac{
		\varepsilon_{\infty,o}\omega_{\mathrm{LO},o}^2
		+
		\left(
		\varepsilon_{2,\mathrm{eff}}+\alpha^2/2
		\right)
		\omega_{\mathrm{TO},o}^2
	}{
		\varepsilon_{\infty,o}
		+
		\varepsilon_{2,\mathrm{eff}}
		+
		\alpha^2/2
	}.
	\label{eq:omega-star-film}
\end{equation}

Table~\ref{tab:thinfilm} shows the resulting surface wave bandwidth fraction $f_{\mathrm{SW}}$ for a 5~nm Bi$_2$Se$_3$ film on various substrates, using the effective evanescent sampling length $D=100~\mathrm{nm}$ in Eq.~(\ref{eq:eps2eff}).

\begin{table}
	\caption{Surface wave bandwidth estimate for a 5~nm Bi$_2$Se$_3$ film on various substrates.}
	\label{tab:thinfilm}
	\begin{ruledtabular}
		\begin{tabular}{lcccc}
			Substrate & $\varepsilon_{\mathrm{sub}}$ & $\varepsilon_{2,\mathrm{eff}}$ & $f_{\mathrm{SW}}$ hBN & $f_{\mathrm{SW}}$ SiC \\
			\hline
			SiO$_2$ & 3.9 & 5.8 & 48\% & 56\% \\
			CaF$_2$ & 6.7 & 8.4 & 39\% & 46\% \\
			Al$_2$O$_3$ & 9.4 & 11.0 & 33\% & 40\% \\
			Bulk Bi$_2$Se$_3$ & --- & 41.0 & 11\% & 15\% \\
		\end{tabular}
	\end{ruledtabular}
\end{table}

\begin{figure}
	\centering
	\includegraphics[width=1.0\linewidth]{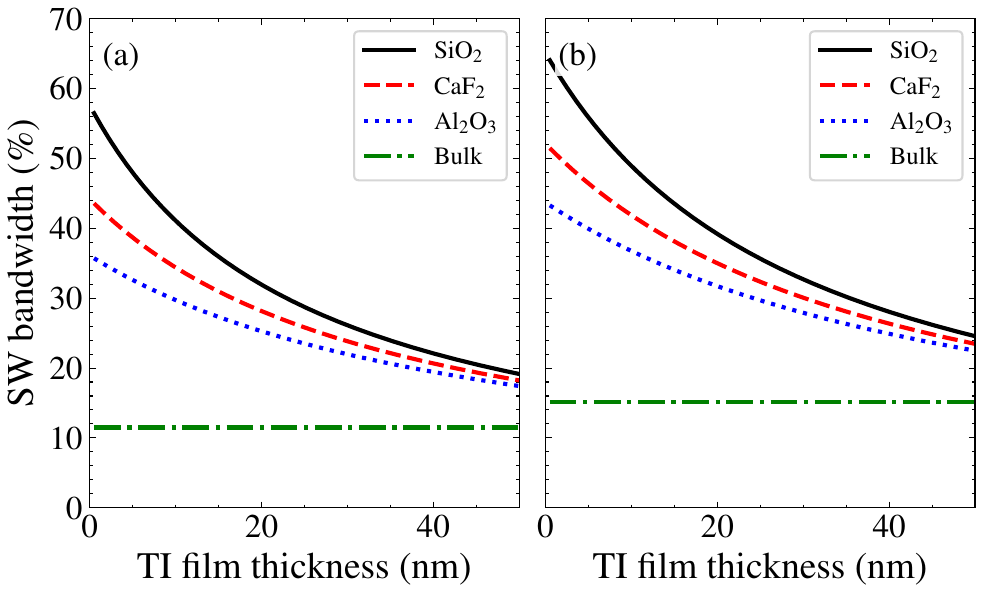}
	\caption{Surface wave bandwidth fraction $f_{\mathrm{SW}}$ as a function of Bi$_2$Se$_3$ film thickness for thin-film TI configurations on different dielectric substrates. Panels show two hyperbolic partners: (a) hBN and (b) SiC. The curves correspond to SiO$_2$, CaF$_2$, and Al$_2$O$_3$ substrates, while the horizontal dash-dotted line shows the corresponding bulk Bi$_2$Se$_3$ limit.}
	\label{fig:thinfilm-bw}
\end{figure}

\begin{figure}
	\centering
	\includegraphics[width=1.0\linewidth]{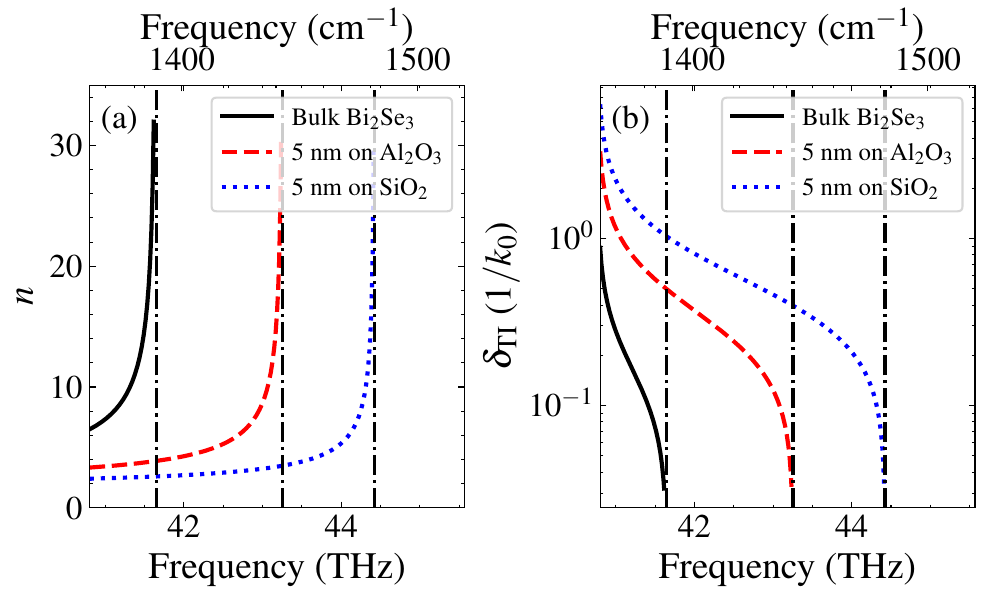}
	\caption{Dispersion and confinement for bulk and thin-film TI configurations at the hBN/Bi$_2$Se$_3$ interface. (a) Effective index $n=\beta/k_0$ for bulk Bi$_2$Se$_3$ and for 5 nm Bi$_2$Se$_3$ films on Al$_2$O$_3$ and SiO$_2$ substrates. (b) Corresponding TI-side penetration depth $\delta_{\mathrm{TI}}$, shown in units of $1/k_0$. The vertical dash-dotted lines mark the corresponding $\omega_*$, the upper boundaries of the allowed surface wave windows.}
	\label{fig:thinfilm-disp}
\end{figure}

The effective medium estimate predicts a substantial broadening of the surface wave window. For example, a 5~nm Bi$_2$Se$_3$ film on SiO$_2$ gives $f_{\mathrm{SW}}=48\%$ for hBN, which is more than four times larger than the corresponding bulk Bi$_2$Se$_3$ value. Figure~\ref{fig:thinfilm-bw} shows the dependence of $f_{\mathrm{SW}}$ on the Bi$_2$Se$_3$ film thickness for both hBN and SiC. Figure~\ref{fig:thinfilm-disp} compares the resulting dispersion curves and TI-side penetration depths for bulk Bi$_2$Se$_3$, Bi$_2$Se$_3$ on sapphire, and Bi$_2$Se$_3$ on SiO$_2$.

The trend in Figs.~\ref{fig:thinfilm-bw} and~\ref{fig:thinfilm-disp} follows directly from the replacement $\varepsilon_2\to\varepsilon_{2,\mathrm{eff}}$ in the band gap condition. Lowering $\varepsilon_{2,\mathrm{eff}}$ shifts $\omega_*$ to higher frequencies within the Reststrahlen band and therefore increases the allowed interval for the surface wave. Among the configurations in Table~\ref{tab:thinfilm}, the largest bandwidth is obtained for SiC with a 5~nm Bi$_2$Se$_3$ film on SiO$_2$, for which $f_{\mathrm{SW}}=56\%$.

These thin film values should be interpreted as design estimates. A quantitative treatment of ultrathin films would require a multilayer calculation that includes the finite thickness of the TI, the substrate response, and possible hybridization of the two topological surface states.

\subsection{Lossy case, artificial hyperbolic material}

\begin{figure}
	\centering
	\includegraphics[width=1.0\linewidth]{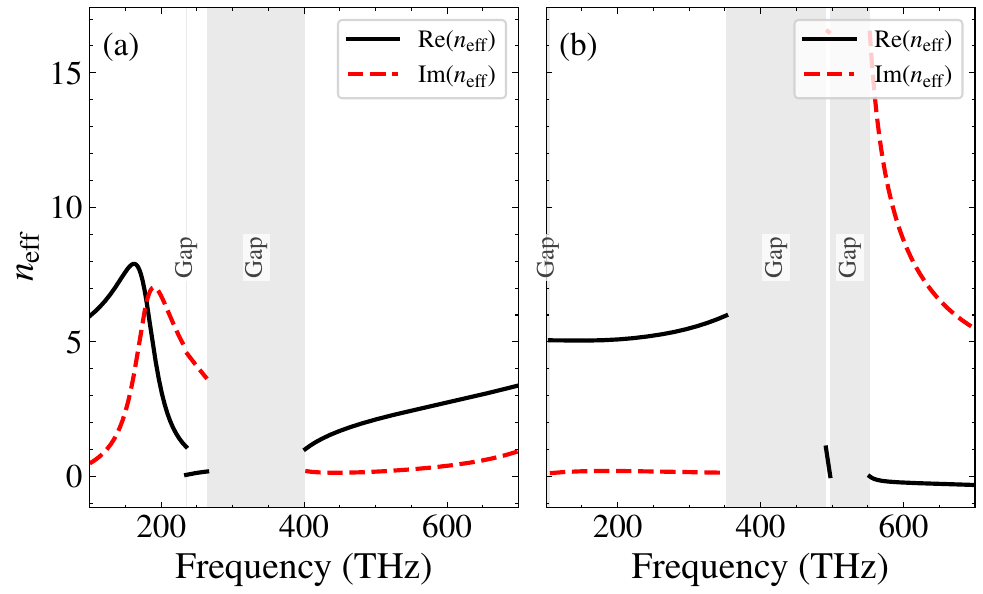}
	\caption{Complex effective index $n_{\mathrm{eff}}$ of the dissipative artificial Ti-Si/Bi$_2$Se$_3$ effective medium surface wave. Panels show the two optical axis orientations: (a) tangential axis, $\mathbf l=\mathbf t_y$, and (b) normal axis, $\mathbf l=\mathbf n$. The black solid and red dashed curves show $\operatorname{Re} n_{\mathrm{eff}}$ and $\operatorname{Im} n_{\mathrm{eff}}$, respectively. Shaded regions mark gaps in the retained physical branch, where the complex branch-selection conditions are not simultaneously satisfied.}
	\label{fig:lossy-neff}
\end{figure}

Figure~\ref{fig:lossy-neff} compares the complex effective index for the dissipative Ti-Si/Bi$_2$Se$_3$ effective medium surface wave for two optical axis orientations. For the tangential orientation, $\mathbf l=\mathbf t_y$, the real part varies strongly across the branch and the imaginary part becomes comparable to it over part of the interval, indicating a strongly attenuated lossy plasmon-polariton branch. For the normal orientation, $\mathbf l=\mathbf n$, the branch is smoother and the imaginary part remains much smaller. Thus, optical axis orientation is not a minor geometric detail: for lossy artificial hyperbolic media, it can determine whether the surface mode is strongly damped or remains relatively well guided.

\begin{figure*}
	\centering
	\includegraphics[width=0.8\linewidth]{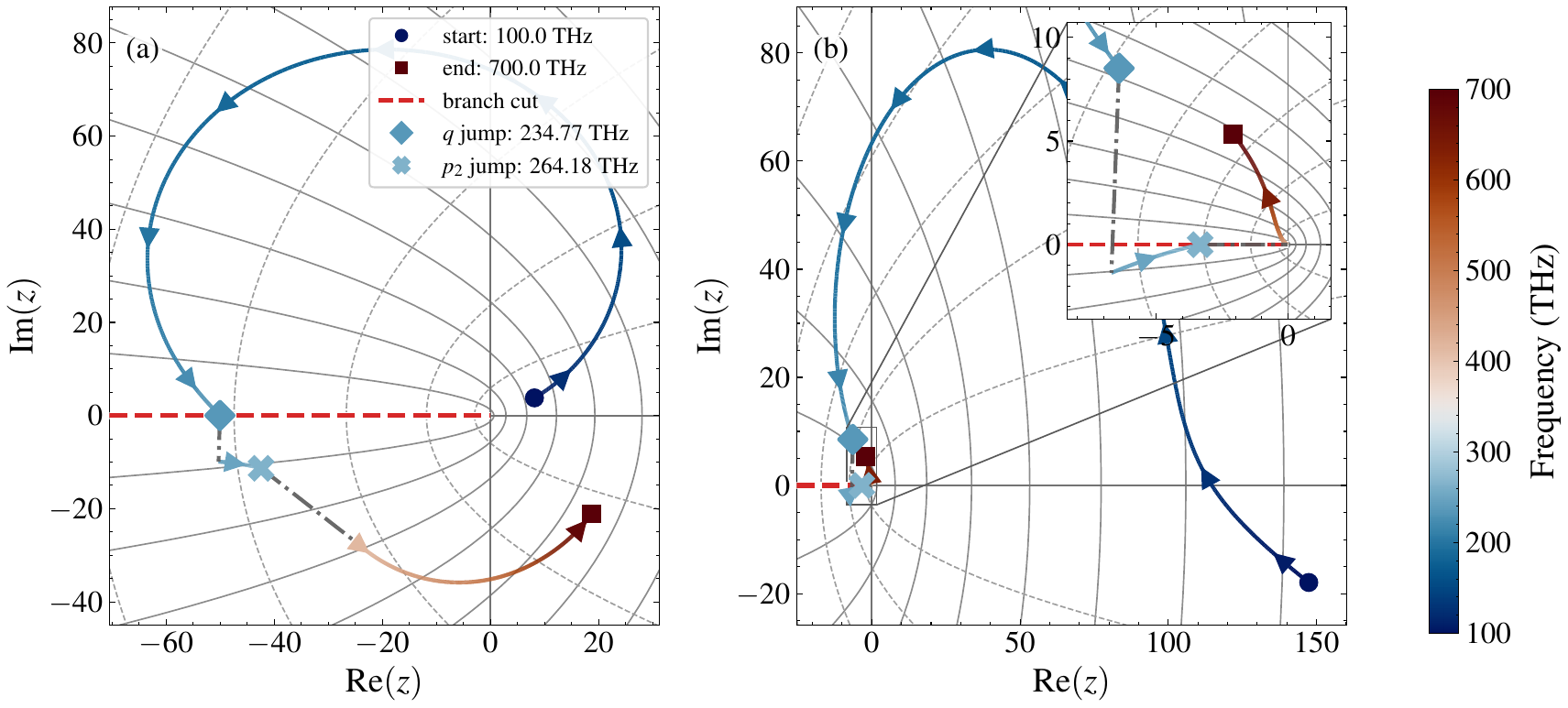}
	\caption{Representative complex plane representation of the square root arguments for the tangential axis case, $\mathbf l=\mathbf t_y$. The plotted trajectories are (a) $z_q=n_{\mathrm{eff}}^2-\varepsilon_2$, associated with the TI-side decay constant $q=k_0\sqrt{z_q}$, and (b) $z_{p_2}=n_{\mathrm{eff}}^2-\varepsilon_o$, associated with the hyperbolic medium TM decay constant $p_2=k_0\sqrt{z_{p_2}}$. Color indicates frequency, and arrowheads indicate increasing frequency. The dashed red line denotes the square root branch cut. Markers identify the start and end of the tracked branch and the frequencies at which the trajectories reach the branch cut; crossing or approaching this cut changes the selected square root branch and produces gaps in the effective-index curve.}
	\label{fig:complex-planes}
\end{figure*}

As discussed in Sec.~\ref{sec:bandgap-lossy}, apparent discontinuities in a lossy case can arise when the complex square root arguments approach the branch cut of the square root. Figure~\ref{fig:complex-planes} illustrates this mechanism for the tangential case, $\mathbf l=\mathbf t_y$. The same branch selection procedure applies to the normal axis configuration. Panel~(a) shows the trajectory of $z_q=n_{\mathrm{eff}}^2-\varepsilon_2$, while panel~(b) shows the trajectory of $z_{p_2}=n_{\mathrm{eff}}^2-\varepsilon_o$, both tracked as the frequency increases from $100$ to $700$~THz. The first interruption of the retained branch in Fig.~\ref{fig:lossy-neff}(a) occurs when $z_q$ reaches the square root branch cut at approximately $234.77$~THz. At that point the root defining $q=\sqrt{z_q}$ cannot be continued on the same sheet while preserving the physical condition $\operatorname{Re}q>0$, so the admissible branch must be reselected and the tracked $n_{\mathrm{eff}}$ branch develops the first gap. Although $z_{p_2}$ has not yet reached its own branch cut, panel~(b) shows that it also undergoes a simultaneous jump because the underlying solution $n_{\mathrm{eff}}$ has switched branches. The second interruption occurs when $z_{p_2}$ itself reaches the branch cut at approximately $264.18$~THz, see the inset in Fig.~\ref{fig:complex-planes}(b). There the square root branch for $p_2=\sqrt{z_{p_2}}$ must be changed in order to keep $\operatorname{Re}p_2>0$, which produces the second gap in Fig.~\ref{fig:lossy-neff}(a); correspondingly, $z_q$ shows a companion jump in panel~(a). Thus each gray gap in Fig.~\ref{fig:lossy-neff}(a) is tied to one of the two square root arguments touching the branch cut, while the other argument exhibits a simultaneous jump because both are determined by the same dispersion root.

\begin{figure}
	\centering
	\includegraphics[width=1.0\linewidth]{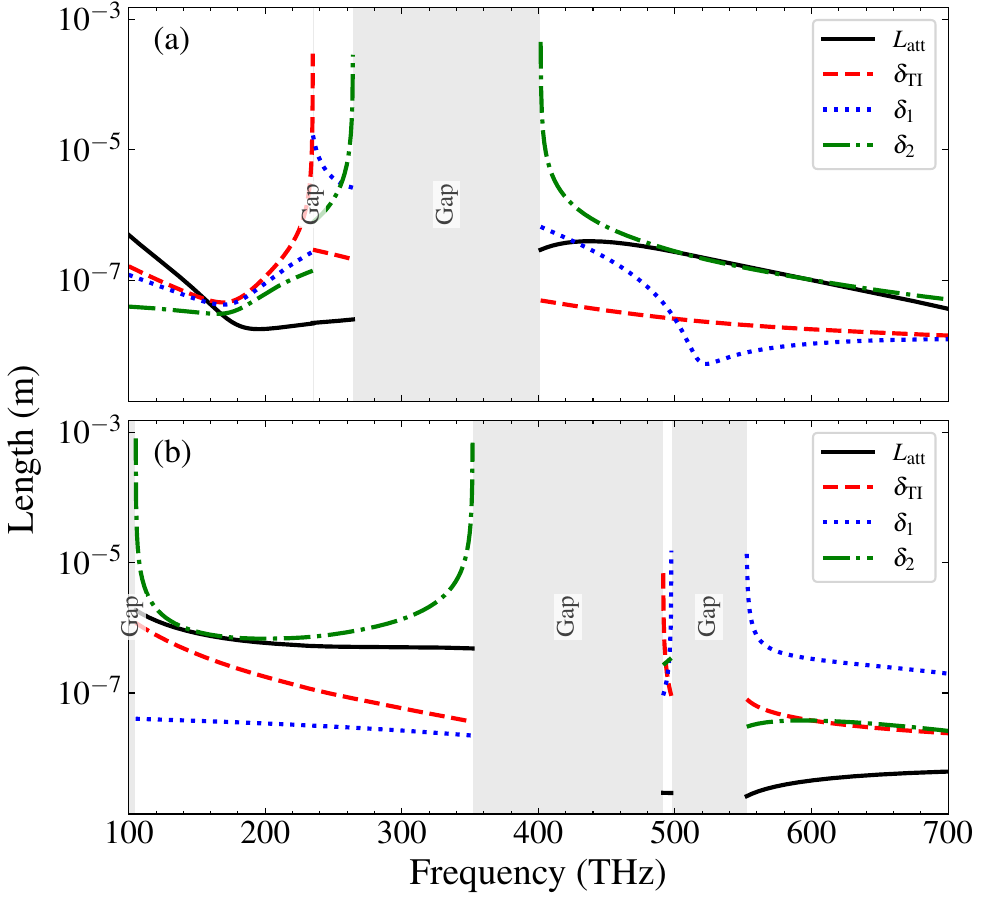}
	\caption{Characteristic length scales for the dissipative artificial Ti-Si/Bi$_2$Se$_3$ effective medium surface wave. The black solid curve shows the intensity attenuation length $L_{\mathrm{att}}$ along the propagation direction, while the red dashed, blue dotted, and green dash-dotted curves show the transverse penetration depths $\delta_{\mathrm{TI}}$, $\delta_1$, and $\delta_2$, respectively. Panels correspond to the two optical axis orientations: (a) tangential axis, $\mathbf l=\mathbf t_y$, and (b) normal axis, $\mathbf l=\mathbf n$. Shaded regions mark gaps in the retained physical branch, where the complex branch-selection conditions are not simultaneously satisfied.}
	\label{fig:lossy-attenuation}
\end{figure}

The attenuation length is shown in Fig.~\ref{fig:lossy-attenuation}. For $\mathbf l=\mathbf t_y$, $L_{\mathrm{att}}$ decreases sharply in the lossy part of the branch, so propagation over distances much larger than the modal size is not expected there. For $\mathbf l=\mathbf n$, the attenuation length is larger and varies more smoothly. Thus, the normal-axis geometry provides the more robust surface-guiding configuration for the case considered here.

The penetration-depth analysis leads to the same conclusion: large decay lengths occur near branch edges, where one of the localization constants approaches zero and the mode becomes weakly confined. For the normal-axis geometry, this behavior is the lossy continuation of the lossless localized interval $\varepsilon_2<n^2<\varepsilon_e$. In the dissipative problem, $n_{\mathrm{eff}}^2$ is complex, so localization is imposed through $\operatorname{Re}q>0$, $\operatorname{Re}p_1>0$, and $\operatorname{Re}p_2>0$, rather than through an inequality between complex quantities.

\begin{figure}
	\centering
	\includegraphics[width=1.0\linewidth]{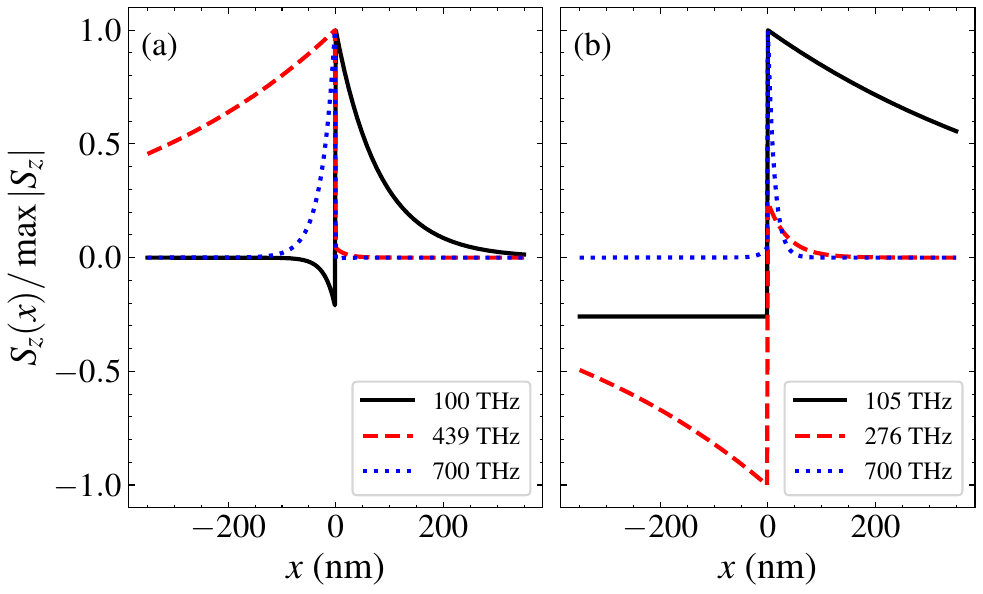}
	\caption{Normalized longitudinal Poynting-density profiles $S_z(x)/\max |S_z|$ in Ti-Si/Bi$_2$Se$_3$ medium surface wave. Panels show the two optical axis orientations: (a) tangential axis, $\mathbf l=\mathbf t_y$, and (b) normal axis, $\mathbf l=\mathbf n$.}
	\label{fig:lossy-poynting}
\end{figure}

Finally, Fig.~\ref{fig:lossy-poynting} shows representative normalized profiles of the longitudinal time averaged Poynting density,
\begin{equation}
	S_z(x)
	=
	\frac{c}{8\pi}
	\operatorname{Re}\left(E_xH_y^*-E_yH_x^*\right).
	\label{eq:poynting-main}
\end{equation}
The profiles show that the energy flow can have opposite signs in the hyperbolic medium and in the TI. This behavior is typical for surface waves involving negative or anisotropic permittivity components: the total guided power is determined by the balance between forward and backward energy flow on the two sides of the interface. Therefore, for lossy case, attenuation length, penetration depth, and Poynting-flow distribution must be considered together when assessing experimental observability.

\section{Discussion}
\label{sec:discussion}

The dispersion relations in Eqs.~(\ref{eq:disp1}) and~(\ref{eq:disp2}) were first obtained in~\cite{lyashko2017surface} for general material permittivities. Here, this formulation is developed into explicit existence and localization criteria for realistic interfaces between hyperbolic materials and topological insulators. In the lossless case, these criteria yield analytical expressions for the allowed frequency interval, the surface wave bandwidth, and the penetration depths. In the dissipative case, they provide a framework for determining which solutions of the complex dispersion equation correspond to physically localized and attenuated surface waves.

The surface waves considered here are related to several known classes of interface modes, but their confinement mechanism is distinct. In contrast to conventional Dyakonov waves at anisotropic and isotropic interfaces~\cite{d1988new}, the present modes involve hybridization of TE and TM field components through the topological boundary conditions. In contrast to Dirac plasmons in topological insulators~\cite{di2013observation}, the negative permittivity required for confinement is supplied by the hyperbolic material rather than by free surface carriers. Compared with axionic surface polaritons~\cite{qi2014surface}, the hyperbolic material introduces an additional anisotropic decay structure, which restricts the frequencies at which localized modes can exist.

For lossless natural hyperbolic materials, the allowed frequency interval is governed primarily by the dielectric contrast across the interface. The topological coupling adds a shift of order $\alpha^2$ to this dielectric band-edge condition, but the shift is negligible on the frequency scale of the material dispersion considered here. The large permittivity of bulk Bi$_2$Se$_3$ shifts the surface wave band edge toward the transverse optical phonon frequency and leaves only a narrow spectral window, as reflected in the bandwidths listed in Table~\ref{tab:bandgap}. Reducing the effective dielectric response on the TI side, for example by placing a thin TI film on a lower permittivity substrate, moves this boundary to higher frequencies within the Reststrahlen band and widens the surface wave interval. The analytical criteria also show that frequency variation changes both the propagation constant and the modal confinement through the penetration depths.

The dissipative titanium and silicon effective medium interfaced with Bi$_2$Se$_3$ illustrates the additional structure that appears when the propagation constant and transverse decay constants are complex. In this regime, roots of the dispersion equation must be interpreted together with the decay behavior of the fields in both media. The tangential axis configuration provides a representative example: the admissible solution is interrupted at frequencies where the complex arguments of the square roots defining the transverse decay constants cross their branch cuts in the complex plane. The same type of analysis applies to the normal axis configuration, although the resulting effective index branch is smoother for the parameters considered here. This behavior reflects the analytic structure of the complex decay constants and must be accounted for when identifying localized modes in dissipative media.

The lossy calculations further show that the orientation of the optical axis has a direct effect on the admissible surface wave solutions. For the tangential axis configuration, the imaginary part of the effective index increases substantially over part of the solution, and the attenuation length can become comparable to the modal size. For the normal axis configuration, the corresponding effective index curve is smoother and exhibits weaker attenuation for the same effective medium parameters. These results indicate that, in dissipative structures, the localization of the field, the propagation loss, and the energy flow profile must be evaluated together when determining whether a solution of the dispersion equation represents an observable guided surface wave.

The axion term enters the boundary conditions linearly and therefore produces a TE-TM field admixture of order $\alpha$, whereas its correction to the dispersion relation and the band edge is of order $\alpha^2$. Consequently, measurements of the effective index or band-edge position are intrinsically insensitive probes of the bulk topological response for conventional topological-insulator parameters, while polarization-resolved measurements may provide a more direct signature. More broadly, the calculation establishes a quantitative baseline: any substantially larger change in dispersion or confinement should be attributed to dielectric dispersion, geometry, loss, or additional surface conductivity rather than to the bulk axion term. This separation of scales identifies the mechanisms that control practical tuning and provides a reference for structures designed to enhance the topological response.

The frequency range and modal confinement predicted for the natural material interfaces are compatible with near field optical characterization. The calculated surface waves lie in the mid-infrared range, approximately 7.2--7.4~$\mu$m for hBN and 12.0--12.2~$\mu$m for $\alpha$-MoO$_3$. These wavelengths are accessible to scanning near field optical microscopy, which has been widely used to image phonon polaritons in related materials~\cite{dai2014tunable,ma2018plane}. The calculated effective indices, $n\sim$~6--30, correspond to substantial wavelength compression, while the hybridization of TE and TM field components suggests that polarization resolved near field measurements could help isolate the axion induced contribution.

\section{Conclusion}
\label{sec:conclusion}

We have derived analytical criteria for surface electromagnetic waves at interfaces between hyperbolic materials and topological insulators and applied them to both lossless natural materials and dissipative effective media. In the lossless model, the surface wave band edge satisfies $|\varepsilon_o(\omega_*)|=\varepsilon_2+\alpha^2/2$. The dielectric contrast therefore provides the dominant contribution to the band gap, while the bulk axion coupling produces a much smaller correction of order $\alpha^2$. The same coupling also weakly modifies the dispersion and penetration depths within the allowed interval. An important outcome is therefore the quantitative separation of dielectric and topological effects: dielectric properties control practical tuning of the surface wave spectrum, while the bulk topological term establishes a perturbative baseline and is more naturally sought through polarization mixing than through a measurable shift of the band edge.

For hBN/Bi$_2$Se$_3$ and $\alpha$-MoO$_3$/Bi$_2$Se$_3$, the large mid-infrared permittivity of bulk Bi$_2$Se$_3$ produces narrow surface wave windows. A thin TI film on a lower permittivity substrate can substantially broaden these windows by reducing the effective dielectric response.

For dissipative hyperbolic media, solving the dispersion equation is not sufficient to establish the existence of a guided surface wave. The titanium and silicon effective medium interfaced with Bi$_2$Se$_3$ shows that, when the propagation constant and transverse decay constants are complex, some roots of the dispersion relation do not correspond to fields that are localized near the interface and attenuate consistently during propagation. In this regime, gaps or jumps in the admissible surface wave solution can occur when the complex arguments of the square roots that define the transverse decay constants cross their branch cuts in the complex plane. These interruptions are distinct from band gaps associated with the lossless material response. The physically relevant solution must be identified by requiring transverse decay away from the interface, attenuation along the propagation direction, finite penetration depths, and a consistent energy flow profile.

Overall, the results provide design rules for controlling mid-infrared surface waves at interfaces between hyperbolic materials and topological insulators through dielectric contrast, film geometry, optical axis orientation, and possible magnetic surface responses.

\begin{acknowledgments}
	This paper is dedicated to the memory of Andrei Ivanovich Maimistov (1951--2022), whose profound contributions to nonlinear optics and electromagnetic wave theory, together with his direct involvement in this work, helped shape its scientific foundation. The authors sincerely thank Evgenii Narimanov for valuable discussions, advice, and constructive comments during the preparation of this paper. They also gratefully acknowledge Aleksander Volkov for his assistance in identifying relevant literature and in collecting and refining the physical parameters of the materials considered in this study.
\end{acknowledgments}

\section*{Data Availability}
Data underlying the results presented in this work are available from the corresponding author upon reasonable request.

\bibliography{refs}

\end{document}